\documentclass[epjST]{svjour}
\usepackage{graphicx}
\begin{document}
\title{Extended event driven molecular dynamics for simulating dense granular
matter}
\author{J. Sebasti\'an Gonz\'alez\inst{1,2} \and Dino Risso \inst{3} \and
Rodrigo Soto\inst{1}}

\institute{Departamento de F\'{\i}sica, FCFM, Universidad de Chile, 
Santiago, Chile. \and
Faculty of Engineering Technology, TS, MSM, University of Twente, 
Enschede, Netherlands. \and
Departamento de F\'{\i}sica, Universidad del  B\'{\i}o-B\'{\i}o, 
Concepci\'on,  Chile.}

\abstract{
A new numerical method is presented to efficiently simulate the inelastic hard
sphere (IHS) model for granular media, when
fluid and frozen regions coexist in the presence of gravity. The IHS model is
extended by allowing particles to change their
dynamics into either a frozen state or back to the normal collisional state,
while computing the dynamics only for the particles 
in the normal state. Careful criteria, local in time and space, are designed
such that particles become frozen only at 
mechanically stable positions. 
The homogeneous
deposition over a static surface and the dynamics of a rotating drum are 
studied as test cases. The simulations agree with previous experimental
results. The model is much more efficient than the usual event driven
method and allows to overcome some of the difficulties of the standard IHS
model, such as the existence of a static limit.  
} 
\maketitle

\section{Introduction}
Historically, hard sphere models have been very important in
understanding some physical processes. This is  because they allow for a direct
comparison between the different theoretical approaches and precise
numerical simulations \cite{Allen}.  For example, hard sphere models
of atoms helped in understanding the development
of short range
correlations in liquids and the different roles played by the
repulsive and attractive components of the interatomic potential
energy \cite{Hansen}. Similarly, poly-disperse hard sphere particles
helped to improve
understanding of the jamming transition in glasses \cite{TorquatoBook}.

In a simplified, but extremely useful description of granular matter,
grains are considered as hard spheres. The energy dissipation
through collisions
is included via a restitution coefficient, constituting the
so-called inelastic hard sphere (IHS).  In its simplest form, the
restitution coefficient is constant and affects only the normal
relative velocities, hence modeling soft spheres. More realistic
collisions, including tangential restitution coefficients, static and
dynamic friction, and rotation have been considered \cite{Jenkins}. The
IHS model, in its different forms, has been very successful on
describing the main features of granular flows and it has become a useful
tool to improve understanding of the different phenomena that appear in
granular
matter (see, for example \cite{Cambell90,Jaeger96,Kudrolli04}).

Hard sphere models, in general, are numerically simulated by using the
event driven molecular dynamics (EDMD) method. This method consists in
repeating the following procedure: first to predict the next
collision event between two
particles or with the walls, then to analytically bring forward the system to
that instant, and finally
compute the velocities of the particles after the collision. That is,
the system advances from event to event. This method is very precise
because there is no time discretization of the integration of
trajectories or collisions, as the particle dynamics can be solved analytically
between collisions. It is also efficient because the effective time
step between events per particle matches the mean free time.  Furthermore,
after several
optimizations it takes a time ${\cal O}{(\log N)}$ to attend every
event, where $N$ is the number of particles~\cite{Marin}.

It was soon realized that the simulation of IHS particles
presented the {\em inelastic collapse} problem
\cite{McNamara}: inelasticity increases the probability of
recollisions and it is possible that few of the grains perform an
infinite number of collisions among themselves in a finite
physical time. Obviously, infinite events must be attended
taking an infinite amount of computational time. 
This phenomenon is an artifact of the IHS model in which collisions are
instantaneous while in Nature collisions take a small, but finite time.
This artifact is corrected by regularizing the model: inelastic collisions are
turned into elastic
ones in the case where the relative velocity is small \cite{McNamara} or when
the collision frequency goes above a certain threshold~\cite{TCmodel}.
However, when simulating systems where dense regions coexist with
fluid-like regions under the presence of gravity 
the regularized IHS models becomes very inefficient. In 
dense regions, due to the regularization, particles remain colliding
elastically with small relative velocities and high frequencies, forming
solid-like structures. All these 
collisions produce almost no dynamic evolution, but nevertheless 
constitute the overwhelming majority of all collisional events in the 
system.

Commonly, to overcome this difficulty, the simulation of dense
flows is performed using time driven molecular dynamics of continuous force
models like the Hertzian contacts, the spring-dashpot model, or the ratchet
model \cite{Poschel}. However, realistic values of the Young modulus
need the use of extremely small time steps, making the simulations
extremely costly. Sometimes, to speed up simulations, smaller values of the
Young modulus are used. Other approach is the method of
contact dynamics, in which the static reaction forces between grains
are obtained as a solution of a large set of
equations at a considerable computational cost \cite{Moreau}. 

In this article we present an extension to the IHS
model of granular systems, that remains efficient in situations where
solid-like and fluid regions coexist in presence of gravity and confining walls.
It has been reported \cite{TCmodel} that
the IHS model has three major drawbacks: 1) The number of
collisions can diverge, i.e. the {\em inelastic collapse} can occur; 2)
all interactions are binary, multiparticle contacts cannot occur;  and 3) no
static limit exists. Besides there are no persistent contacts, as happens
in dense flows. Our model proposes a solution to these problems 
except for the second one, which is only considered partially because binary collisions 
is indeed one of the key ingredients that allows to perform efficient simulations.

\section{Simulation strategy} \label{sec.algorithm} 
\begin{figure}[htb]
\centering\includegraphics[angle=0,width=0.8\columnwidth]{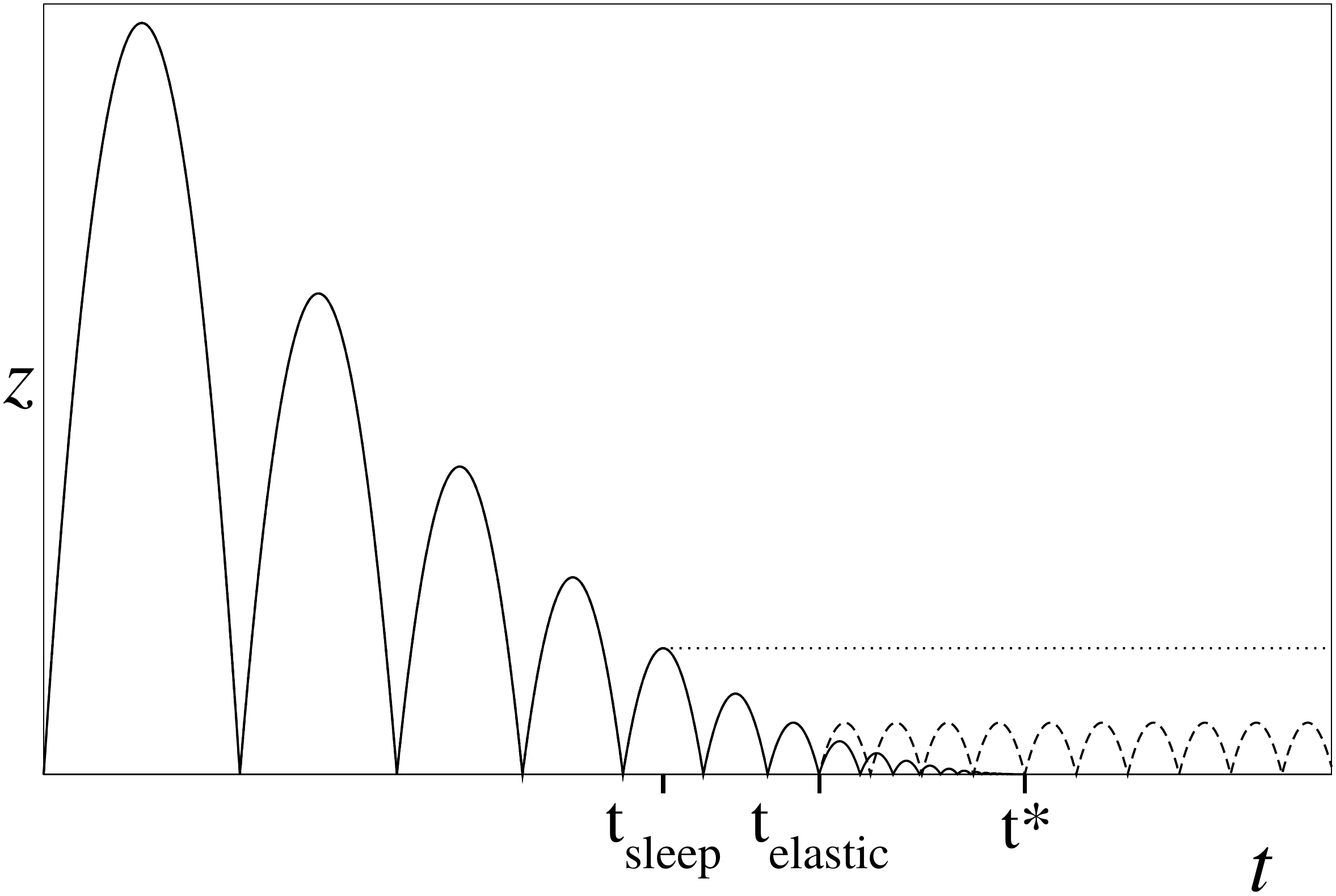}
\caption{Height of a particle bouncing inelastically against a static
wall. With the full inelastic dynamics the particle stops after infinite
collisions at $t^*$ (solid line). If, at $t_{\rm elastic}$, collisions are turn
into elastic for small velocities,  the particle continues bouncing
indefinitely (dashed line). Our model {\em sleeps} the particle if the
velocity is small enough at $t_{\rm sleep}$ and after that it remains motionless
(dotted line). The sleeping time $t_{\rm sleep}$ is located in a slightly later time than the instant in which $V<V_{\rm sleep}$ 
such that the particle is not exactly at contact with the wall.}
\label{fig.bouncing}
\end{figure}

Consider the
simple case of a particle bouncing inelastically against a static
wall. If $\alpha$ is the restitution coefficient with the wall, $V_0$
is the velocity just before the first bounce, and $g$ is the gravity,
the particle would stop at a time 
$t^*=2V_0/(1- \alpha )g$
after an
infinite number of collisions~\cite{viscoelastic}.
If, to avoid this
inelastic collapse, the model is regularized such that collisions were set to elastic when velocities
become smaller that $V_c$, the particle would remain colliding very
frequently for an infinite time, with an irrelevant
dynamics. The solution we propose is that, when the velocity reaches
$V_{\rm sleep}>V_c$, the particle is {\em asleep} and becomes frozen. In this
state the particle does not move and does not accelerate either; physically, this is possible because the particle is supported on the wall. 
To avoid sleeping a particle at exactly contact with
another, which would lead to ill defined dynamics when awake, it is
asleep in an advanced position (halfway to the next collision that it
would have had if it continues with the normal dynamics). Figure~\ref{fig.bouncing} summarizes the described collision sequences.

Therefore, for granular flows, we generalize the state of the
particles to indicate the type of dynamics they follow: normal or
frozen. Normal particles have the regularized IHS dynamics, while frozen
particles do not move or accelerate either with respect to a fixed ground.
For the sake of simplicity we will call {\em sleep} to the pass of
a particle from normal to frozen state, and {\em wake up} to the contrary.
To simulate this extension to the IHS model we will use the EDMD method where
sleeping and waking up particles will be treated as events as well.
The key point is to find efficient, as well as physically sensible,
criteria to sleep or wake up particles. These criteria should be local in time
and space to maintain the efficiency of the event driven algorithm. 

A particle is asleep if its velocity is smaller than a threshold
value $V_{\rm sleep}$ and it is in a mechanically stable position. To
check
this last condition, without making a detailed force analysis, which
is anyway ill defined in hard sphere models, we propose a
heuristic method. A particle approaching a mechanically stable position
proceeds with a series of collisions with other particles where, at
least one of them, is already a frozen particle. When approaching this
equilibrium position the velocity must be decreasing but, to avoid to
erroneously
identify a particle that is climbing over another, it is also demanded
that the particle is going down. In summary a particle is
asleep if when colliding with a frozen particle: (a) it is going down and
(b) its velocity is smaller than at the previous collision and smaller
than $V_{\rm sleep}$. 
Sleeping particles partially solves one of drawbacks of the IHS model: frozen particles have multiple contacts
but normal ones only have binary interactions.

\begin{figure}[htb]
\centering\includegraphics[angle=0,width=0.8\columnwidth]{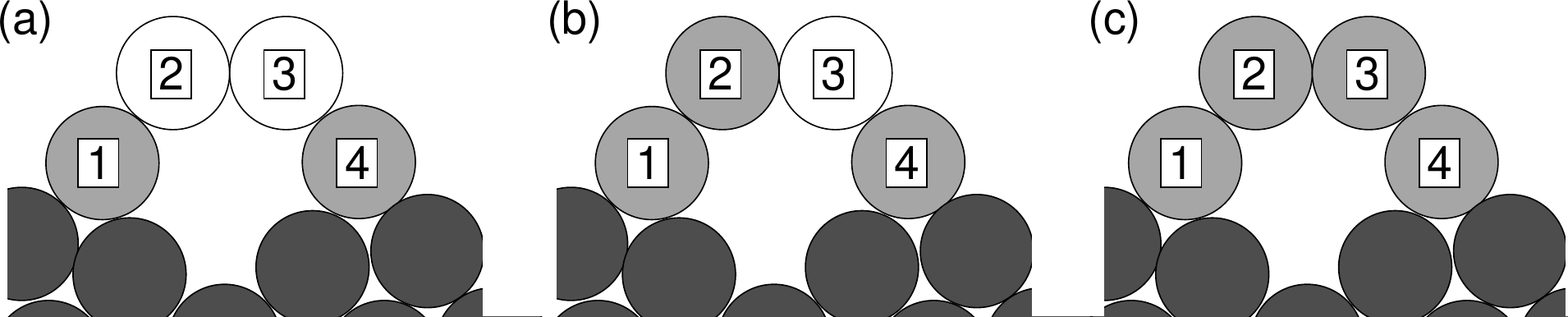}
\caption{Dynamic sequence in two dimensions allowing to build up arches of
frozen particles. In (a) normal particles 2 and 3 have velocities converging to
zero through a series of frequent collisions 
between themselves and with particles 1 and 4, which are frozen and
already supported over others; in (b) 2 sleeps after a collision with 1; finally in (c)
3 sleeps after a collision with 2 or 4. The arch can also be
built by sleeping first 3 and then 2.}
\label{fig.arches}
\end{figure}

Note that when sleeping a particle, it is not necessary that all particles
below it are already frozen but just one. With this approach it is
possible to build up arches as represented in Fig. \ref{fig.arches}, a
fundamental object in granular
solids. This possibility is ruled out in other algorithms like the bottom-to-top
\cite{Visscher}.  

In Fig. \ref{fig.arches}, particle 2 is asleep in (b) fulfilling all
conditions, which is only possible if particle 3 is converging to equilibrium
as well. But using of a finite value for $V_{\rm sleep}$ could consider
particle 3 to be converging while it is in a marginally unstable position.
That is, a particle could be erroneously asleep 
on top of another in a marginally unstable position although its velocity would
have seemed to converge to zero. However, this problem is solved by  
the self-check protocol described later.

Intuitively, to wake up a frozen particle, it must be hit by another particle
with high enough velocity. However, as the frozen particle is supported over
others, the 
collision process is complex and some heuristic must be taken. For simplicity 
a frozen particle is {\em awake} if it is hit with a velocity 
larger than $V_{\rm wake up}$; otherwise the collision is performed as if the 
mass of the frozen particle is infinite and there is a restitution coefficient
$\alpha_{\rm 
frozen}<\alpha$, which models in a simplified way the multiple collisions inside
the frozen region~\cite{restsolid}. Below, it is shown that, besides the 
collisional method to wake up a particle, they must be self-checked and waken up
spontaneously if they loose their mechanical stability.

Sleeping a particle is an inelastic process because its 
energy, although small, is lost. If there is a sequence of repeated sleeping and 
waking up processes, an inelastic collapse can occur. This sequence can take
place when a particle that is on top of a frozen region is collided frequently
from above as in avalanches or in a deposition over a surface.
To avoid this inelastic collapse, the energy 
that the particle had when  it was asleep is stored for a time of 
the order of $2V_{\rm sleep}/(1-\alpha_{\rm frozen})g$, that is, the inelastic
collapse time over a frozen region. 
If it is awake before
that time this energy is reinjected. 

A note should be done regarding the prediction 
of the collision time
between grains, that is obtained from the equation $|\vec r_1(t)-\vec
r_2(t)|=d$. The relative distance between two normal grains does
not depend on gravity and the equation is quadratic, which is simply
solved. But in the case of a normal and a frozen particle
the equation becomes quartic and good root finding methods
should be used to avoid missing solutions, i.e. collisions.
In the case of a normal particle that just collided with a frozen one, the
prediction of the recollision time reduces to a cubic equation because one
solution is already known (the present time).

The particles diameter $d$ and the gravitational acceleration $g$ 
define natural microscopic units for time and velocity, $\sqrt{d/g}$ and
$\sqrt{gd}$,
respectively. The threshold velocities must be small compared with 
this velocity scale.  
To avoid the inelastic collapse in the bulk we  set $V_c=10^{-4}\sqrt{gd}$.

In order to test the model in demanding regimes, we have considered quite inelastic systems where the interparticle restitution coefficient is
fixed to $\alpha=0.7$. In absence of a more refined model for the particle-solid restitution coefficient its value is fixed arbitrarily to a small value
$\alpha_{\rm frozen}=0.4$.

\section{Homogeneous deposition}\label{sec.deposition}
First, we consider the homogeneous
deposition of grains in a container to obtain a granular packing. 
Different methods have been used in the literature for that purpose: 
(a) In the the bottom-to-top reconstruction method, particles are added 
one-by-one waiting to the previous one to have found an equilibrium 
position~\cite{Visscher}.  This method is unable to form arches and therefore it cannot reproduce all the 
phenomenology of granular 
solids. Besides, the kinetic energy of the particles is lost and it is not
possible for a particle to wake up others. 
(b) The simulation of grains with dissipative Hertz contacts needs 
small time steps to resolve collisions. Hence, it becomes extremely time
consuming when realistic values of the Young modulus are used. 
(c) In Ref.~\cite{torquato} grains grow until a jamming situation
is attained, representing a different physical mechanism.

Here, we let $N=4800$ particles to fall down in an homogeneous rain at
a small volume fraction ($\phi\simeq 0.04$) in a box of size $10d\times
10d\times 500d$.  Each particle falls at a constant vertical velocity
$V_{\rm fall}\simeq -2.5\sqrt{gd}$, until it has its first collision. From then on 
the particle adopts the  dynamics described in the
previous section. This is analog to a deposition with a moving heap keeping
constant the distance between the heap and the free
surface~\cite{exp_deposition}.
The base is composed of a random packing of frozen grains that
never wake up, in order to avoid a local crystalline order in the lower layers. 
Periodic boundary conditions
are used in the lateral directions. Simulations stop  when all particles are  frozen
with zero kinetic energy.

\begin{figure}[htb]
\centering\includegraphics[width=.9\columnwidth]{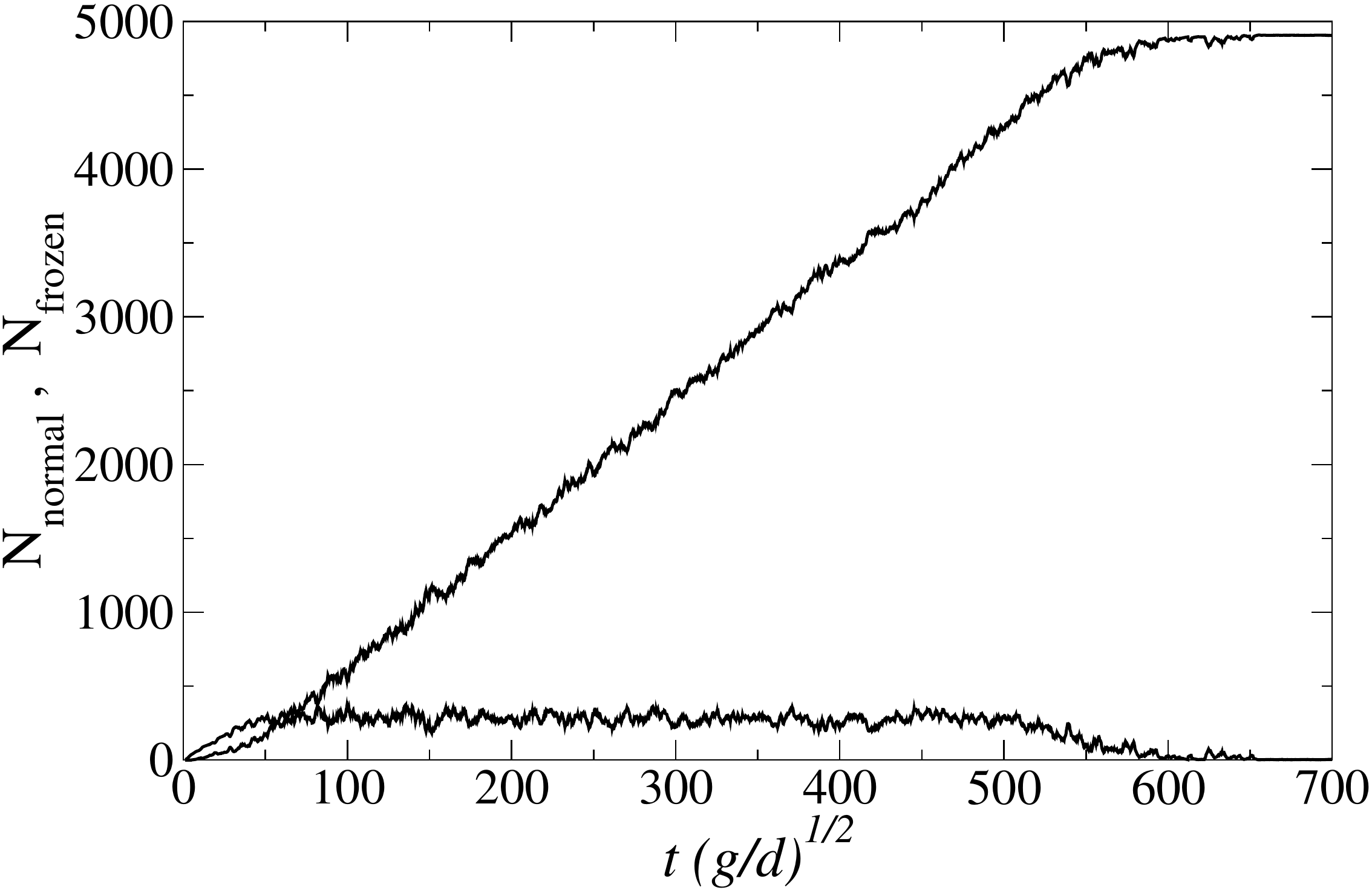} 
\caption{Evolution of the number of normal (lower curve) and frozen (upper
curve) particles during an homogeneous deposition. Particles with constant speed
are not included in the graph.}
\label{fig.Numbers}
\end{figure}

\begin{figure}[htb]
\centering
\centering\includegraphics[width=.75\linewidth]{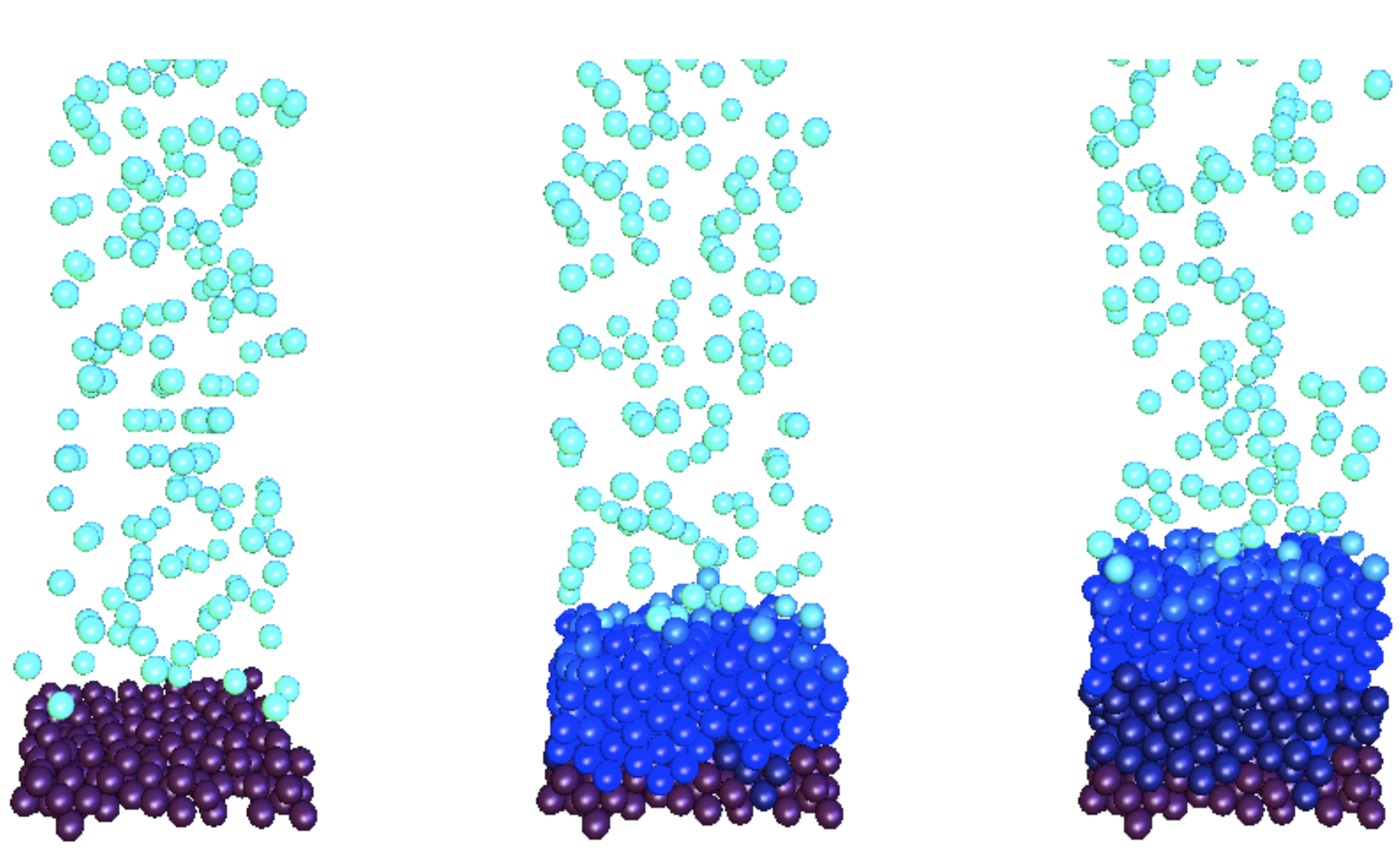}
\caption{Three snapshots of the deposition process for $V_{\rm sleep}=V_{\rm
wakeup}=0.01\sqrt{gd}$ at different instants: left $t=0\sqrt{d/g}$, center
$t=50\sqrt{d/g}$, and right $t=100\sqrt{d/g}$. The base particles that cannot
wakeup are presented in purple, particles in dark blue are frozen,  the particles of the homogenous rain are
presented in light blue, and the normal particles are presented in different
levels of blue depending on their kinetic energy.}
\label{fig.deposicion}
\end{figure}

In Fig.~\ref{fig.Numbers} we present the evolution of the number of normal
and frozen grains in a simulation made with $V_{\rm sleep}=V_{\rm wake
up}=0.01\sqrt{gd}$. The number of normal particles, which are
the effectively simulated ones, only occupy a small region on top of the
deposit and it remains roughly constant as the deposited layer grows (see Fig.
\ref{fig.deposicion}). 
The average number of normal particles is $\langle N_{\rm normal}\rangle =
280\pm2$. Considering a packing fraction equal to that of the random close
packing, RCP, $\phi_{\rm RCP}=0.64$, this corresponds to a
fluidized layer of $4.4 d$ in height.

\begin{figure}[htb]
\centering
\centering\includegraphics[width=.75\linewidth]{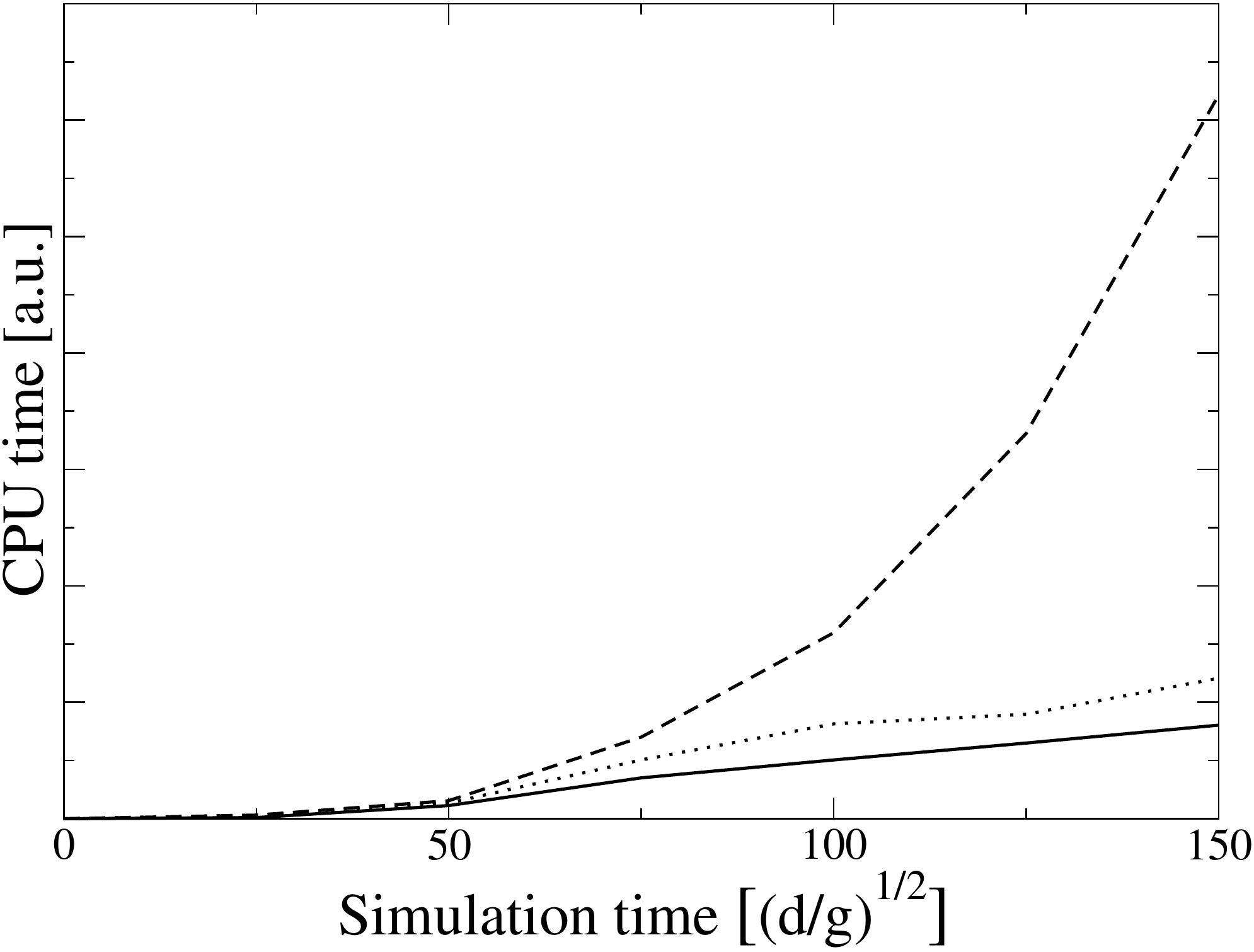}
\caption{CPU time, in arbitrary units, to achieve different simulation
times for three values of the thresholds $V_t=V_{\rm sleep}=V_{\rm
wakeup}$. $V_t=0.0$ (dashed line),  $V_t=0.05\sqrt{gd}$ (dotted line), and 
$V_t=0.10\sqrt{gd}$ (solid line).}
\label{fig.cputime}
\end{figure}

Figure \ref{fig.cputime} presents the CPU time needed to achieve different
simulation times for three values of the threshold velocities with $V_{\rm
sleep}=V_{\rm wakeup}$. It is seen that when then threshold velocities are
finite, the CPU time grows linearly with the simulation time as the normal
particles remain constant through all the simulation as seen in Fig.
\ref{fig.Numbers} although the proportionality constant depends on the value of
the threshold velocities as more particles are normal when they have smaller thresholds. In
the case of vanishing thresholds the CPU time grows at least quadratically 
with the simulation time as the number of normal particles continuously
increases.

A series of simulations were run varying the threshold velocities. 
For simplicity, we choose $V_{\rm sleep}=V_{\rm wake up}=V_t$ in the range 
$0.01\leq V_t/\sqrt{gd}\leq0. 1$. 
The packing fraction, computed in the middle region of the deposit, increases
slightly when the threshold velocities become smaller and it is well fitted by
the expression $ \phi = (0.640\pm0.001) - (0.11\pm 0.01) V_t/\sqrt{gd}$. 
The asymptotic packing fraction
with the full dynamics  is $\phi_0=0.640$, equal to the random close packing value
$\phi_{\rm RCP}=0.64$~\cite{torquato}
and largely above the random loose packing value
$\phi_{\rm RLP}=0.555$~\cite{Oneda}. 
The RCP limit could be attained due to the upper layer that
remains in a fluid state, allowing to efficiently accommodate the grains thanks
to the energy transfer from the deposited grains to the ones that were frozen.
On the contrary, if the fluidization layer is avoided, for example, by
depositing particles one-by-one and setting $V_{\rm wakeup} = \infty$ and
$V_{\rm sleep} = 0.05\sqrt{gd}$ (like in the bottom-to-top algorithm
\cite{Visscher}), the  
packing fraction is much lower: $\phi_{\rm onebyone}=0.586$. 

When simulating packings, a balance should be assumed when choosing the
threshold velocities in order to obtain accurate values for the packing
fraction at an affordable computational cost.  In our runs the total number
of events performed until the simulation stops diverges as $N_{\rm events}
\sim
V_t^{-1.2}$. It should be noted that the packing fraction dependence is much
softer than the computational cost, and therefore in normal applications
high enough threshold values can be taken.

\section{Rotating drum and the self-check protocol}\label{sec.rotatingdrum}
In the case of the simple deposition described above, it is not
crucial the chosen value of $V_{\rm wakeup}$  because the final state
is motionless. A well studied regime where particles change
continuously from normal to frozen, and where we can study the effect of the
wake-up threshold in the dynamics, is the rotating drum \cite{GRD}.

\begin{figure}[htb]
\centering\includegraphics[width=.8\columnwidth]{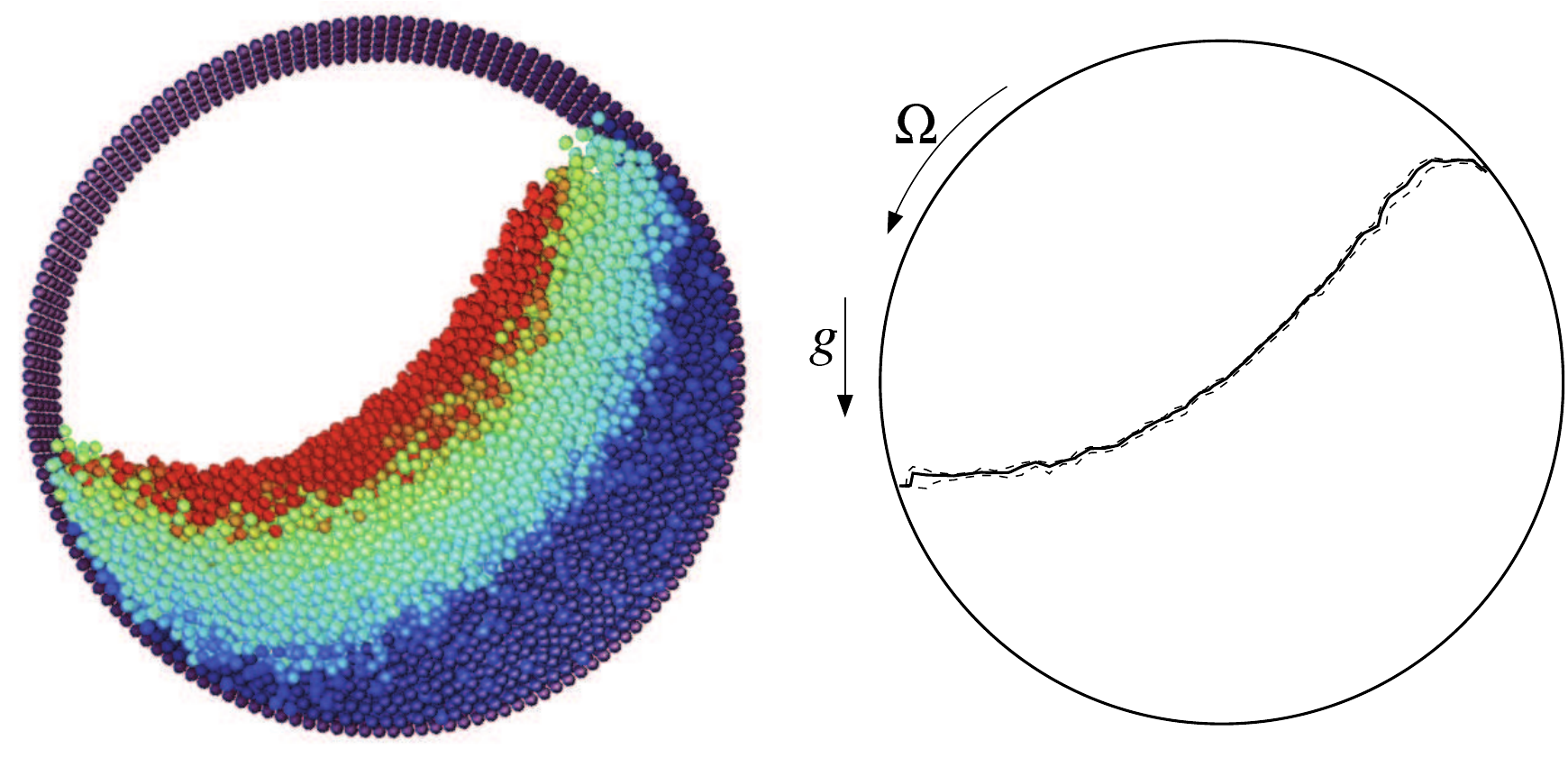}
\caption{Left: Snapshot of the rotating drum in the stationary state. Colors
indicate the kinetic energy. Right: Average packing fraction iso-lines. The
solid line is for $\phi=0.5\phi_{RCP}$ and the dashed lines are for 
$\phi=0.25\phi_{RCP}$ and $\phi=0.75\phi_{RCP}$.}
\label{fig.snapshotdrum}
\end{figure}

We consider a half-filled cylinder of radius
$R=25d$, thickness $L=5d$, with periodic boundary conditions
in this direction, and fixed angular frequency $\Omega=0.075 \sqrt{g/d}$.
The rotation of the drum is modeled by changing the direction of the
gravitational force at a fixed rate. Doing this, we can emulate moving walls
without actually have to move them.  In the case of moving the walls, the
velocity thresholds would refer to the relative
velocity of the colliding particles and not to the absolute velocities.
This approach neglects the centrifugal
forces, that are small for the angular velocity we use. 
The drum
wall is made of a regular array of frozen particles that cannot wake up, with 
a transverse and arc separation of $1.05d$ and $1.0d$, respectively (see 
Fig.~\ref{fig.snapshotdrum}).

Avalanches form in the rotating drum because particles that are raised to the
top loose their mechanical stability. However, the collisional wake up mechanism
described in section \ref{sec.algorithm} is not able to wake them up
under this condition. To overcome this problem, a self-check protocol is set:
each particle is checked every $\tau_{\rm check}$ to verify that all the 
particles that were supporting it when it became frozen are still there.
According to Fig. \ref{fig.arches}, a particle is supported over neighbor
particles (closer than $1.05d$) that are frozen below it (defined in terms of
the direction of gravity) and eventually of normal type above it. Note that, a
change in the direction of gravity could turn a mechanically stable
configuration into an unstable one. If the test fails the particle is waken up;
if it is still in a mechanically stable position it will sleep again after a
few collisions, otherwise it will continue in the normal state.
The self-check mechanism allows also to wake-up particles that were erroneously
asleep on top of others in marginally unstable positions. 

In the
simulation of the deposition, $\tau_{\rm check}$ was set to infinity
because in that setup, no particle will loose its mechanical
stability once settled down. In the present case, a small value of $\tau_{\rm
check}$ would lead to
too many particles being checked, increasing the computational cost and
eventually to an inelastic collapse;  a large
value of it would artificially maintain particles in mechanically
unstable positions. As a compromise we have chosen $\tau_{\rm check}=0.1\sqrt{d/g}$. 

The particles that are waken up after the self-check and continue to be
normal will avalanche through the inclined free surface, waking up other
particles through collisions. Figure \ref{fig.snapshotdrum} presents a typical
snapshot of
the system once the system has reached the stationary state.  Some 
packing fraction iso-lines, measured by averaging different snapshots of two compete rotations of the drum, are presented in the
same figure. A clear free surface develops with an abrupt density drop.

\begin{figure}[htbp]
\centering\includegraphics[width=.9\columnwidth]{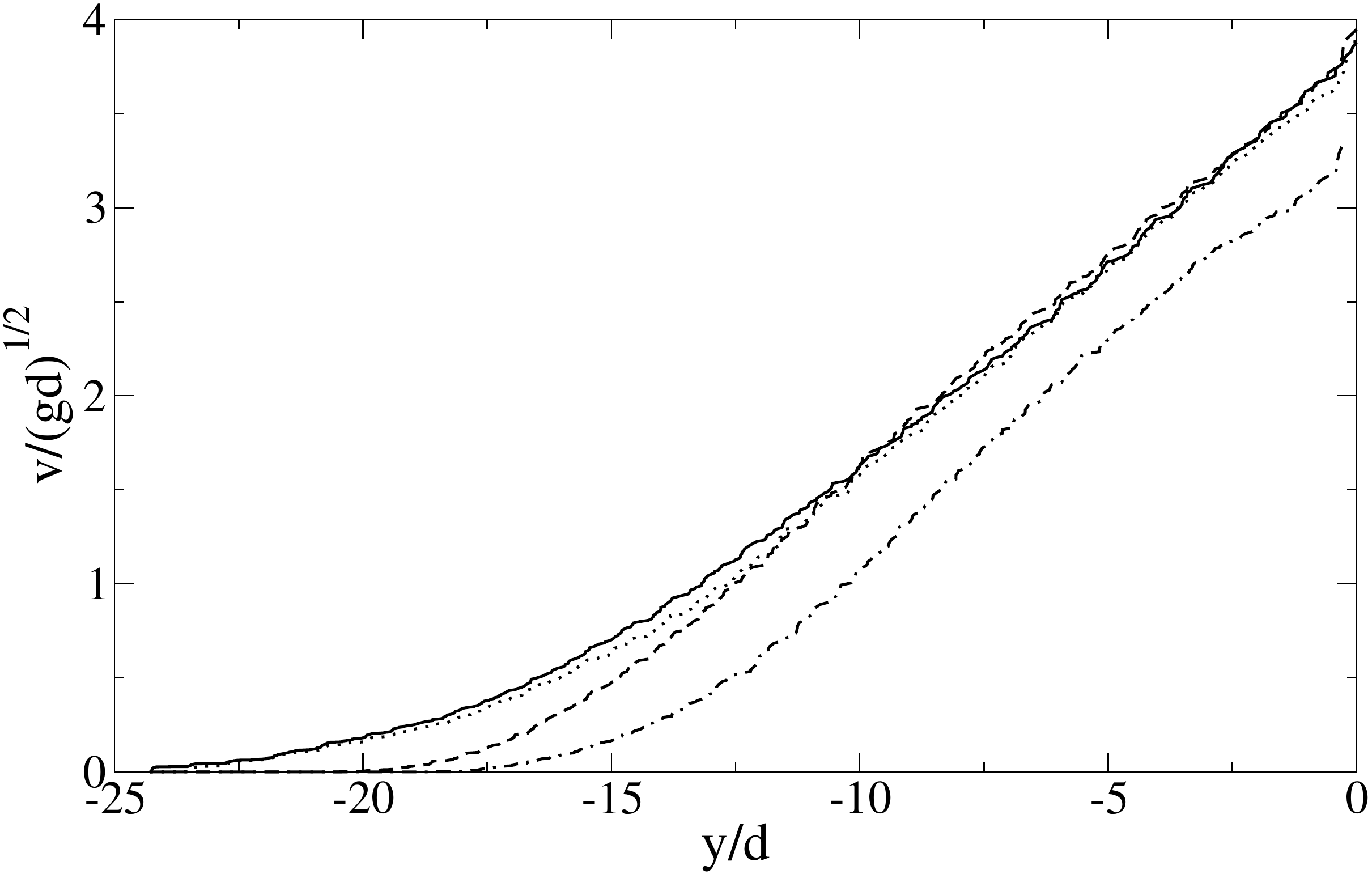}
\caption{Velocity profile in the reference frame of the rotating drum
for different values of the wake-up threshold: $V_{\rm wake up}=0.01\sqrt{gd}$
(solid
line), $V_{\rm wake up}=0.05\sqrt{gd}$ (dotted line), $V_{\rm wake
up}=0.25\sqrt{gd}$ (dashed
line), $V_{\rm wake up}=0.50\sqrt{gd}$ (dot-dashed line).
Distances are measured in a thin slab, perpendicular to the free surface in the
center of the system, with the origin placed at the free surface.}
\label{fig.velprofile}
\end{figure}

In order to show the effect of $V_{\rm wake up}$ in the dynamics, we present
in Fig. \ref{fig.velprofile} the velocity profiles in the reference frame of
the rotating drum, for fixed $V_{\rm sleep}=0.05\sqrt{gd}$ and different
values of $V_{\rm wake up}$. These profiles should be compared with the
experimental results reported in \cite{GRD}, which show a linear shear band
followed by a frozen region that is static with respect to the drum. In
between, there is a small transition region, where velocity decays
exponentially.

When the wake up threshold is $0.50\sqrt{gd}$ (equivalent to waking
a particle up if it is hit by another particle that fell from a height of
$d/8$) the velocity profile resembles
those found experimentally~\cite{GRD}. 
The system develops a continuous avalanche of a thickness of approximately two thirds of
the drum radius and the rest of the grains remain motionless with respect to the
drum. Due to our resolution, we cannot affirm that
the transition region has an exponential decay. 

However, when the wake up
threshold is too small ($0.05\sqrt{gd}$), the particles near the drum wall
get overfluidized and there is no frozen region. In fact, because of the
regularization, particles near the bottom do
not sleep because they have not reached this threshold but collisions are
already elastic. Therefore they remain with a finite amount of energy
and the bottom region is in an artificial elastic regime. 
This result shows that the IHS model
is inaccurate to describe dense granular regions since the regularization
method, introduced to avoid the inelastic collapse, creates
artificially overfluidized elastic regions. Jamming and freezing is not
possible and solid like regions become softer.
Our model
corrects this by providing an additional
method to dissipate energy --freezing  the particles-- allowing for a better
description of dense flows near a fixed boundary.
Further research, however, is needed to determine criteria to fix $V_{wakeup}$.

\section{Conclusions}\label{sec.conclusions} 
We have extended the IHS model for granular media to
include static particles that have lost all its kinetic energy,
allowing to simulate granular flows coexisting with solid-like
regions. It allows to reach the static limit,
recreating multiple persisting contacts. The model is efficient because collision events are
only computed for particles that are moving, which are the only
dynamically relevant. A combination of strategies to sleep, wake up,
and self-check particles allows to dynamically move particles to frozen
and normal states back and forth.
The model has been used to study two cases that present difficulties
when studied using the standard or regularized IHS model: the homogeneous
deposition until all particles are motionless and the continuous avalanche that develops in a rotating drum.
Our model correctly describes these cases with results that agree with those found experimentally.
Besides, sleeping the particles provides with an additional energy dissipation
mechanism, that is important to describe correctly flows near a fixed boundary. 

The model uses the parameter $V_c$ that has been widely discussed in the literature and three new parameters: 
$V_{\rm sleep}$, $V_{\rm wake up}$, and $\alpha_{\rm frozen}$. The effect of the two threshold velocities
$V_{\rm sleep}$ and $V_{\rm wake up}$ have been discussed in the two examples. 
Nevertheless, an estimation for $\alpha_{\rm frozen}$ needs a model for the interaction of  a grain with a frozen region, which for 
small velocities is lacking. In absence of more information $\alpha_{\rm frozen}$ was fixed arbitrary. Further research is needed.

We mention that as in the two setups we considered, frozen particles that
cannot wake up can be used to build up complex boundaries with no additional
computational cost. 

We thank E. Clement and P. Cordero for fruitful
discussions and S. P\'erez for critical reading the manuscript.
This research is supported by Fondecyt Grants No. 1061112, No.
1070958, and Fondap Grant No. 11980002.

\end{document}